# Dynamical diversity of pulsating solitons in a fiber laser

Hong-Jie Chen[1], Yan-Jie Tan[1], Jin-gan Long[1], Wei-Cheng Chen[2], Wei-Yi Hong[1], Hu Cui[1], Ai-Ping Luo[1,*], Zhi-Chao Luo[1,*], and Wen-Cheng Xu[1,*]

[1] Guangdong Provincial Key Laboratory of Nanophotonic Functional Materials and Devices & Guangzhou Key Laboratory for Special Fiber Photonic Devices and Applications, South China Normal University, Guangzhou 510006, China

[2] School of Physics and Optoelectronic Engineering, Foshan University, Foshan 528000, China

* Corresponding authors: luoaiping@scnu.edu.cn; zcluo@scnu.edu.cn; xuwch@scnu.edu.cn

**Abstract:** Pulsating behavior is a universal phenomenon in versatile fields. In nonlinear dissipative systems, the solitons could also pulsate under proper conditions and show many interesting dynamics. However, the pulsation dynamics is generally concerned with single soliton case. Herein, by utilizing real-time spectroscopy technique, namely, dispersive Fourier-transform (DFT), we reveal the versatile categories of pulsating solitons in a fiber laser. In particular, the weak to strong explosive behaviors of pulsating soliton, as well as the rogue wave generation during explosions were observed. Moreover, the concept of soliton pulsation was extended to the multi-soliton case. It is found that the simultaneous pulsation of energy, separation and relative phase difference could be observed for solitons inside the molecule, while the pulsations of each individual in multi-soliton bunch could be regular or irregular. These findings would further enrich the ultrafast dynamics of dissipative solitons in nonlinear optical systems.

## 1. Introduction

Compared with the Hamiltonian system, the dissipative system is more complicated in the sense that the dissipative system has the dissipative properties of continuous exchange of energy with the environment. In such a dissipative system, gain and loss play an essential role in the formation of dissipative solitons. Owing to its characteristic of energy exchange, the fiber laser can be regarded as a dissipative system and the pulses formed in such a system can be treated as dissipative solitons [1]. In addition to its excellent ability of generating ultra-short pulses which has a wide range of applications in fields from fundamental physics to industrial purposes, the mode-locked fiber laser is also known as an ideal test bed for investigating complex nonlinear dynamics. So far, various striking nonlinear phenomena have been observed in fiber lasers, such as multi-soliton patterns [2-5], vector soliton [6-9] and dissipative soliton resonance [10-13]. However, due to the lack of advanced measurement technologies, these results were not based on a real-time measurement method, which means that some important information may be ignored since the ultrafast nonlinear phenomena cannot be captured by the conventional measurement instruments.

Till recent years, owing to the development of real-time measurement techniques, dispersive Fourier-transform (DFT) has been regarded as one of the standard real-time measurement techniques for resolving ultrafast dynamics of solitons in laser systems [14]. Through the DFT, the spectrum of the soliton could be mapped into a temporal waveform by using dispersive element with enough group-velocity dispersion. Thus, the ultrafast spectral signals can be captured by a real-time oscilloscope with a high-speed photodetector. The DFT technique opens up the way of experimentally investigating the transient dynamics of solitons in lasers. Indeed, it has been employed to observe complex ultrafast nonlinear phenomena, including soliton explosion [15-19], build-up and transient dynamics of solitons [20-25], pulsating solitons [26-28], as well as internal dynamics and build-up of soliton molecules [29-35].

Particularly, pulsating soliton as a distinctive localized structure in dissipative systems has attracted extensive attention [36-43]. Unlike equal amplitude of pulse train generated from the laser, it is found that the amplitude of the pulsating soliton could evolve periodically or in a more complex way along with the cavity roundtrips and the pulsating soliton exists regardless of the sign of the dispersion. In the early years, the properties of the pulsating solitons could merely be observed by the oscilloscope due to the limitation of the measuring apparatus. That is to say, only the temporal intensity information of the pulsating solitons was obtained [38]. With the development of real-time spectroscopy techniques, i.e., DFT, it is possible to unveil the real-time spectral dynamics of pulsating solitons [26-28]. Generally, the soliton pulsating behavior in fiber laser systems was investigated in the single pulse regime. In fact, theoretically the dissipative optical systems could also support the cases of multi-soliton pulsations. However, the reported dynamical behaviors of the pulsating soliton, in particular the multi-soliton pulsation, are still limited, the underlying dynamics have not been fully revealed. Therefore, it is of great interest to investigate the versatile behaviors of pulsating solitons in a fiber laser.

In this work, we firstly observed, to the best of our knowledge, the dynamical diversity of pulsating solitons in a fiber laser by using the DFT. We revealed the real-time spectral evolution from single-soliton pulsation to multi-soliton pulsation. The explosive behavior during the single-soliton pulsation was observed, where the degree of soliton explosions is from weak to strong, and the rogue wave was generated. Then the pulsating behaviors of the multi-soliton patterns were also investigated. Particularly, the solitons inside the molecule show the synchronous pulsations in energy, separation and relative phase difference, while the energy pulsation of each individual in multi-soliton bunch could be either regular or irregular. The experimental results would enrich the transient dynamics of the pulses in laser community and contribute to further understanding the complex features of pulsating solitons in dissipative optical systems.

## 2. Experimental setup

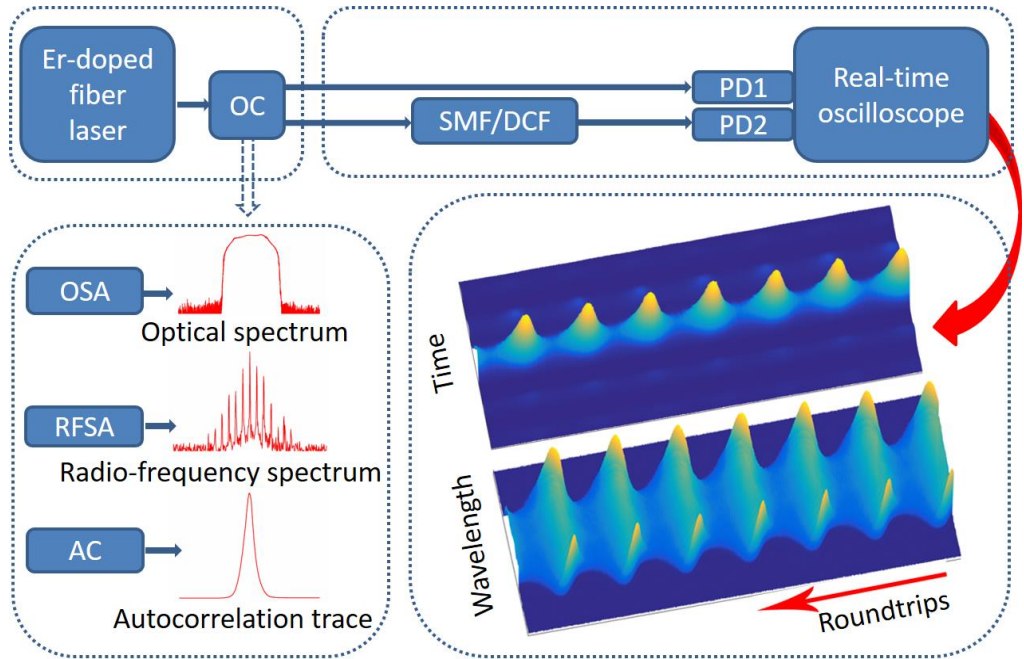

Figure 1. Schematic of the experimental setup.

The dynamical diversity of pulsating solitons was observed in a passively mode-locked fiber laser (See Figure S1 in Supplementary Materials). The schematic of the whole experimental setup is presented in Figure 1. At the laser output, we measure the average optical spectrum and the radio-frequency (RF) spectrum by using an optical spectrum analyzer (OSA, Yokogawa, AQ6375B) and a RF spectrum analyzer (Advantest, R3131A), respectively. The autocorrelation trace of soliton is measured by an autocorrelator (Femtochrome, FR-103WS). Apart from the conventional measurements above, directly observing the spatial-temporal intensity evolution is fulfilled by a real-time oscilloscope (Tektronix, DSA71604B, 16 GHz) with a photodetector (Newport, 818-BB-35F, 12.5 GHz). On the other hand, to measure the spatial-spectral evolution of solitons, the DFT technique is adopted. In order to stretch the temporal solitons and map them into the shot-to-shot spectrum, here for pulsating non-bound-state solitons (the pulsating single-soliton, explosive pulsating soliton and pulsating soliton bunch), a ~16 km long SMF is used, while for the soliton molecule, a piece of dispersion compensation fiber (DCF) with dispersion amount of 2300 ps/nm is adopted. Consequently, combining the photodetector and oscilloscope, the shot-to-shot spectrum can be measured in real-time with resolution of $\Delta\lambda=0.29$ nm and $\Delta\lambda = 0.035$ nm, respectively, for our DFT configuration. It is worth noting that, for the soliton molecules, owing to their separation is quite large, around 106 ps in our experiment comparing to that of soliton pairs in the anomalous dispersion regime, a large total dispersion to stretch the pulsating soliton molecules is therefore in need, which enables two individual corresponding shot-to-shot soliton spectra to overlap completely. In this case, the two superposition spectra can be regarded as a unit, ensuring the accuracy to resolve the shot-to-shot spectral dynamics of pulsating soliton molecules in the spectral domain.

## 3. Experimental results

### 3.1 Pulsating single-soliton and explosive long-period pulsating soliton

#### 3.1.1 Pulsating single-soliton

In our experiment, by adjusting the PC, the dissipative soliton with typical rectangular spectrum can be firstly observed at the pump power of 6.34 mW. But here we do not focus on this type of dissipative soliton which is commonly seen, since the properties of the common dissipative single-soliton operation in fiber lasers, to large degree, have been fully investigated. However, for the solitons with pulsating properties, they have not yet been completely experimentally studied. It is always inspired to explore different patterns of pulsating solitons. Therefore, we are more interested to further reveal the dynamics of pulsating single-soliton, which is also expected to be bunched together as pulsating solitons bunch without soliton interactions or bounded together forming pulsating soliton molecule under the soliton interactions in fiber lasers.

By further finely tuning the PC and adjusting the pump power to 8.43 mW, we were able to observe the pulsating single-soliton. Notably, the average spectrum presented on the OSA has gradient edges instead of conventional rectangular ones, as shown in Figure 2(a). Moreover, the RF spectrum has several satellite peaks around the main peak with separation of 33.8 kHz as depicted in Figure 2(b), indicating the soliton was operating in pulsating regime. Then we performed the real-time measurement techniques to analyze the characteristics of the soliton in both spectral and temporal domains. The 1500 continuous roundtrips evolution of shot-to-shot spectrum reconstructed through DFT is presented in Figure 2(c). One can see that the spectrum breathes periodically. Correspondingly, the soliton energy also varies with the cavity roundtrips with a pulsating period of ~274 roundtrips, corresponding to a modulation frequency of 33.8 kHz in RF spectrum presented in Figure 2(b). Figure 2(d) shows the corresponding temporal evolution of the soliton. It is evident that the width and intensity of the soliton vary periodically with the cavity roundtrips, emerging pulsating properties.

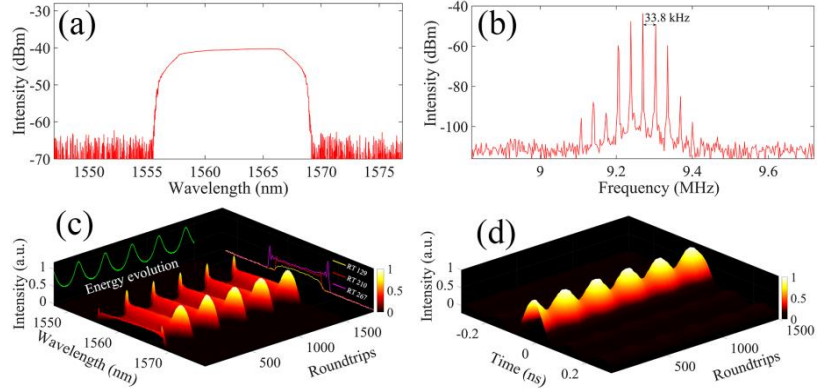

Figure 2. The characteristics of the single-soliton pulsation. (a) The spectrum obtained from the OSA. (b) The RF spectrum. (c) The shot-to-shot spectra and corresponding energy evolution. (d) Pulse train.

Furthermore, we fixed the pump power and subtly adjusted the orientation of the PC to better display the characteristics of pulsating solitons. We find that the spectrum displayed on the OSA varies correspondingly, namely, the spectrum can vary from the relative flat top to the gradient one (See Figure S2 in Supplementary Materials). Moreover, the special spectrum is related to the degree of soliton pulsation parameters (strong or weak pulsation). That is to say, the larger the gradient extent of the spectral edges, the stronger the modulation of pulsating soliton energy and amplitude is. Particularly, when the gradient extent of spectral edges is slight enough, the energy modulation is also very slight, showing a tendency toward a stable mode-locked state. The variation of the spectral gradient extent could be attributed to the change of the pulsating states in the laser cavity. In the experiment, the adjustment of the PC would change the cavity loss, namely, the transmission of the cavity varies, leading to the variation of the pulsating states. Consequently, the characteristics of the mode-locked spectrum would change correspondingly.

### 3.1.2 Explosive long-period pulsating soliton

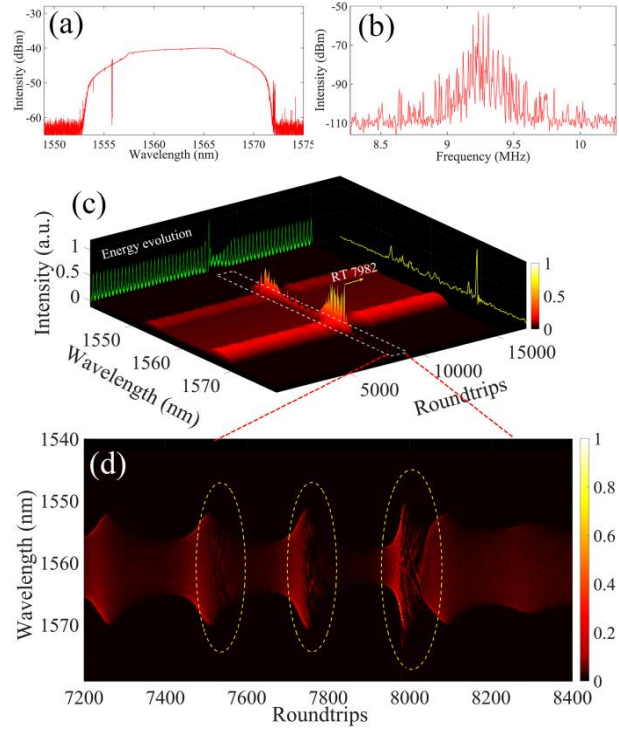

Figure 3. The explosive long-period pulsating soliton operation. (a) The spectrum obtained from the OSA. (b) The RF spectrum. (c) The shot-to-shot spectra and corresponding energy evolution. (d) The enlarged explosive pulsating soliton region.

In the process of exploring the pulsating single-soliton, especially, we also observed that the pulsating single-soliton collapsed intermittently and returned back to the original pulsating state after roundtrips of dozens to around one hundred during the evolution. This soliton nonlinear phenomenon, in fact, is similar to the soliton explosion [15-19]. When adjusting the orientation of the PC, in addition to achieving the regularly pulsating single-soliton at the pump power of 8.98 mW, we noticed a unique average spectrum from the OSA showing a high instability: spikes and dips stochastically occur on the spectrum. One typical spectrum is presented in Figure 3(a). And Figure 3(b) shows the corresponding RF spectrum, which has many irregular peaks. To acquire the possible diverse characteristics of the pulse, we recorded 15000 roundtrips real-time spectra with DFT, as shown in Figure 3(c). Obviously, during the evolution of the pulsating soliton, we can see the explosive pulsating soliton spectrum with extreme intensity peak around the 8000th roundtrip and the corresponding pulse energy suddenly raises. To clearly show the detailed explosive process of pulsating soliton, we take out 1200 roundtrips single-shot spectra from the 15000 roundtrips evolution process, as shown in Figure 3(d). From the 7200th roundtrip to the 8400th roundtrip, three explosive events of the pulsating soliton corresponding to three pulsating periods can be observed, and the explosive degree strengthens gradually. Note that, comparing to the explosions of pulsating soliton around the 7550th and the 7800th roundtrips, the pulsating soliton explosion around the 8000th roundtrip is much more highly chaotic and sustains longer roundtrips (about 100 roundtrips).

For these three different explosion events, the spectrum collapse sustains certain roundtrips and recovers to the regular pulsating soliton. As for the strongest explosive pulsating soliton event, the spectrum broadens dramatically. To illustrate more details, one typical single-shot spectrum is drawn during the explosion at the 7982th roundtrip, as presented in the inset of Figure 3(c). Here, the spectrum collapse can be found clearly, and the explosive spectrum has intensive peaks. Besides, we note that the RF spectrum in Figure 3(b) has quite many satellite peaks, which also shows a slight chaotic feature. Given that the pulsating soliton behavior is along with the explosive events and long-period pulsation (~242 roundtrips) at the same time in our experiment, we regard this unique pulsating soliton pattern as the ***"explosive long-period pulsating soliton"***. Moreover, we investigated the extreme waves in the spectral domain by recording the maximum intensity of the single-shot spectrum. It is found that spectral rouge waves generated in such an explosive long-period pulsating soliton regime, which is shown in the Supplementary Materials (See Figure S3).

## 3.2 Pulsating dual-soliton bunches

### 3.2.1 Irregular pulsating dual-soliton bunch

In the following, before seeking for the pulsating soliton molecules which are expected to be formed in a laser system, we tried to explore the pulsating solitons bunch to see the possible dynamics difference between the single-soliton pulsation and multi-soliton bunch pulsation. More importantly, since the soliton molecule can be treated as a unique pattern of multi-soliton bunch when the separation between the individuals is small enough, it might be instructive to further find the pulsating soliton molecules.

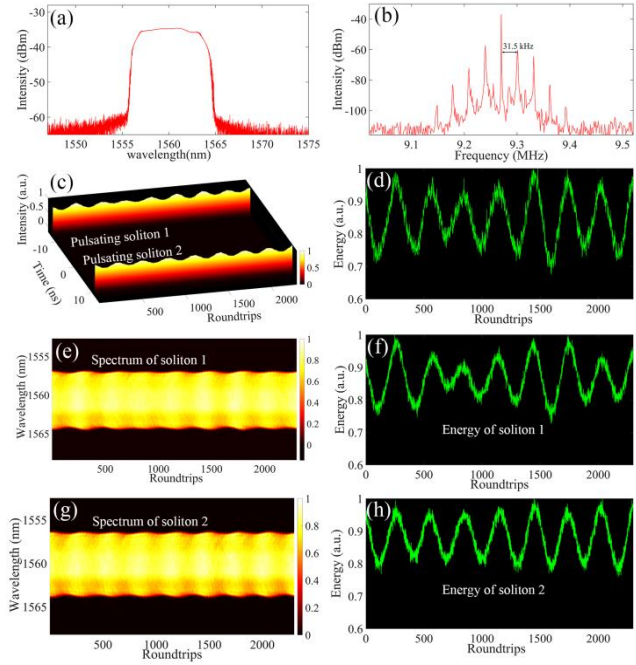

Figure 4. Irregular pulsating dual-soliton bunch operation. (a) The spectrum obtained from the OSA. (b) The RF spectrum. (c) The pulse train. (d) The total energy evolution of the dual-soliton bunch. (e) and (g) The shot-to-shot spectra of the two solitons. (f) and (h) The energy evolution of the two solitons.

As a rule, multi-soliton bunch is expected to be generated under the stronger pump power in the laser system. Therefore, by increasing the pump power to 12.06 mW, the stable mode-locked dual-soliton bunch could firstly be achieved. Likewise, the corresponding spectrum with steep edges was observed. Then, we finely tuned the PC to look for the average optical spectrum with gradient edges on the OSA, as shown in Figure 4 (a). Here, note that the gradient range of the spectral edges is smaller comparing with that in the pulsating single-soliton regime. The corresponding RF spectrum is shown in Figure 4(b), which also has several satellite peaks around the main peak with separation of 31.5 kHz, meaning that the dual-soliton bunch presents pulsating behavior. To further verify the existence of pulsating solitons bunch, temporal evolution of the laser output was measured directly by the real-time oscilloscope and shown in Figure 4(c). It is evident that the pulsating dual-soliton bunch has been observed. To further get insight into the shot-to-shot spectral evolution of the pulsating solitons bunch, via the DFT, the two pulsating solitons were stretched into corresponding real-time spectra. Instead of relatively strong periodic pulsation of spectrum in pulsating single-soliton, both two periodic spectra show a slight pulsation as illustrated in Figures 4(e) and (g). Figures 4(d), (f) and (h) display the corresponding energy evolution of solitons bunch and each individual. Particularly, it can be seen that, during the periodic evolution of soliton energy in dependence of roundtrips, irregular fluctuation happens to the periodic energy pulsation of soliton 1, while the energy of soliton 2 keeps relatively regular pulsation. For this energy pulsation difference, we suppose that one of the solitons experiences a small perturbation, leading to the variation of peak power for this soliton. Thus, the pulsating behavior changes correspondingly. In this case, the different pulsation of the two solitons could be observed. Although irregular energy pulsation takes place to the pulsating dual-soliton bunch, the pulsation frequency of the whole dual-soliton bunch with 31.5 kHz (corresponding to ~294 roundtrips) still can be measured by the RF spectrum analyzer, as shown in Figure 4(b).

### 3.2.2 Dual-soliton bunch with regular pulsation

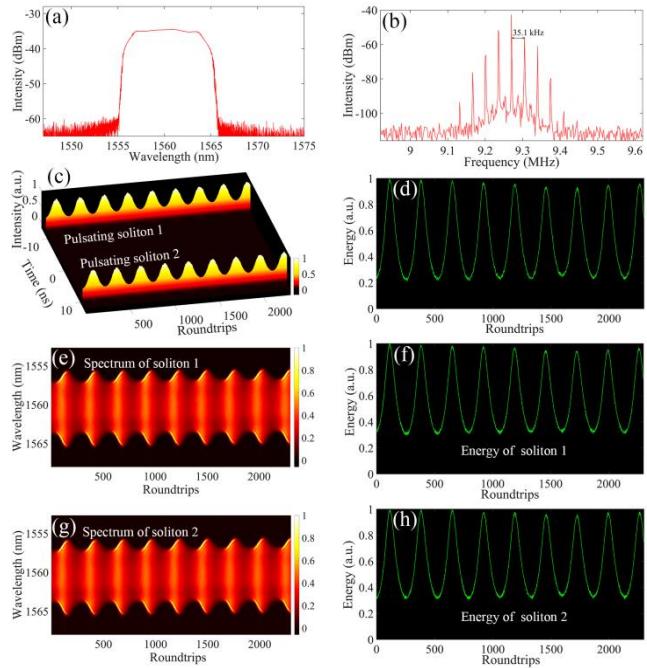

Figure 5. Regular pulsating dual-soliton bunch operation. (a) The spectrum obtained from the OSA. (b) The RF spectrum. (c) The pulse train. (d) The total energy evolution of the dual-soliton bunch. (e) and (g) The shot-to-shot spectra of the two solitons. (f) and (h) The energy evolution of the two solitons.

Then we increased the pump power to 12.63 mW and carefully tuned the PC, dual-soliton bunch with regular pulsation was achieved. The spectrum from the OSA is shown in Figure 5 (a). The periodic evolutions of temporal and spectral intensity are represented in Figures 5(c), (e) and (g), which are further confirmed by a RF spectrum analyzer. Similarly, the RF spectrum in Figure 5(b) displays multiple peaks with frequency difference of ~35.1 kHz, which well agrees with ~264 roundtrips pulsating period of the dual-soliton bunch in Figures

5(c), (e) and (g). From Figures 5(c), (e) and (g), we can clearly see that the pulsations of the dual-soliton bunch are regular and much stronger in both temporal and spectral domains comparing to those of the dual-soliton bunch with irregular pulsation in Figure 4. In addition, due to the increase of pump power, the pulsation period (~264 roundtrips) is also shorter than ~294 roundtrips pulsating period of the dual-soliton bunch with irregular pulsation. In terms of the shot-to-shot spectra, the real-time spectral characteristics are similar to those of the pulsating single-soliton (See Figure S4 in the Supplement Materials). The corresponding energy evolutions of solitons bunch, soliton 1 and soliton 2 are depicted in Figures 5(d), (f) and (h), respectively. Obviously, all three energy pulsating processes are regular and have same period, which means in this case the pulsating dual-soliton complex is fairly stable.

### 3.3 Pulsating soliton molecules

In the following, we aimed for verifying the possible existence of pulsating soliton molecules in the fiber laser which have few reports so far. Actually, soliton molecules can be regarded as a unit when the individual solitons bound together under their interactions. Two particular characteristics of soliton molecules: the spectrum with modulation and the autocorrelation trace with multiple peaks, are useful to identify whether the fiber laser operates in the bound state regime.

Here, based on the characteristics of pulsating solitons we observed above, especially the gradient edges of optical spectrum directly displayed on the OSA, as well as the features of soliton molecules spectrum with interference fringes, we sought for the pulsating soliton molecules. When fixing the pump power at 13.67 mW, we obtained pulsating soliton molecule by controlling the PC, and the results are summarized in Figure 6. Figure 6(a) shows the spatial-spectral dynamical evolution. One can clearly see that the spectra with modulation breathe periodically, namely the spectra experience periodic broadening and compression. For better displaying the interference fringes of the soliton molecule spectrum, we also show the spectra with a small range in Figure 6(d). Correspondingly, the energy of the soliton molecule varies periodically, as presented in the inset of Figure 6(a), showing pulsating property with a period of ~355 roundtrips, which is well consistent with the frequency difference of 26.5 kHz in RF spectrum in Figure 6(c). Here, it should be noted that because the output power is too low to be transmitted in the DCF, the real-time spectral dynamics via DFT was recorded after an erbium-doped fiber amplifier (EDFA). The corresponding temporal evolution of the soliton molecule is further demonstrated in Figure 6(b). From Figure 6(b), one can see that two pulsating solitons bound as a unit and evolves stably. Based on both spectral and temporal characteristics, it demonstrates that pulsating dual-soliton molecule has been observed in our fiber laser. Then we further investigated the temporal separation and relative phase difference between the two solitons inside the molecule from the DFT spectra of the soliton molecule. It is found that the separation and phase difference between the two solitons simultaneously vibrate with the roundtrips, as shown in Figure 6(e). The soliton separation oscillates with amplitude of 1.7 ps and a period of ~355 roundtrips. Meanwhile, the phase difference oscillates with amplitude of $0.34\pi$ and a period of ~355 roundtrips. Figure 6(f) shows the evolution trajectory of the two interacting solitons in the phase plane, indicating that the oscillations possess stably periodic behavior. Thus, we observed the soliton molecule pulsation, which pulsates not only in energy, but also in soliton separation and phase difference.

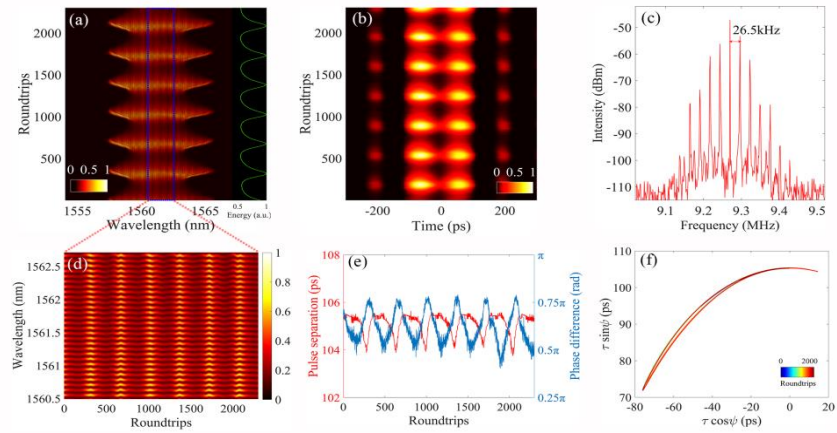

Figure 6. Characteristics of periodic pulsating soliton molecule. (a) The real-time spectra evolution measured with DFT, inset: the energy evolution. (b) The temporal evolution. (c) The RF spectrum. (d) The enlarged spectra. (e) The soliton separation and relative phase difference as a function of the roundtrips. (f) The corresponding interaction planes.

When the PC was further adjusted slightly, the soliton molecule pulsated as a unit could be still observed. That is, the two solitons consisting of the molecule possess synchronous pulsations. However, in this case the soliton molecule did not pulsate in a periodic way. Figure 7(a) shows the real-time spectral dynamics of pulsating soliton molecule with aperiodic behaviors. The details of real-time spectral evolution with shorter spectral range are also given in Figure 7(d). As can be seen here, the interference fringes of mode-locked spectrum present the irregular (or non-perfect periodic) patterns, indicating that the separation and phase difference between the two solitons inside the molecule irregularly evolve with roundtrip time. The whole energy evolution of the pulsating soliton molecule further demonstrated that the pulsating behavior is irregular, as depicted in the inset of Figure 7(a). Correspondingly, the pulse train of the soliton molecule is also shown in Figure 7(b). Here, it can be clearly seen that the two solitons pulsate as a unit, but not in a periodic manner. Moreover, because of the aperiodic pulsation of soliton molecule, the side peaks of RF spectrum with broader bandwidth could be observed under this state, as presented in Figure 7(c). As mentioned above, the interference fringes indicate that the soliton molecule evolves when propagating in the laser cavity. Likewise, to gain insight into the dynamical characteristics of the pulsating soliton molecule from the soliton separation as well as phase difference point of view, we provide the evolution of the soliton separation and phase difference extracted from the DFT spectra in Figure 7(e), where the aperiodic evolving (even random) trend could be observed. Moreover, the corresponding interaction plane between the two solitons is plotted in Figure 7(f). It further proves that the pulsating behavior of the soliton molecule is not stable.

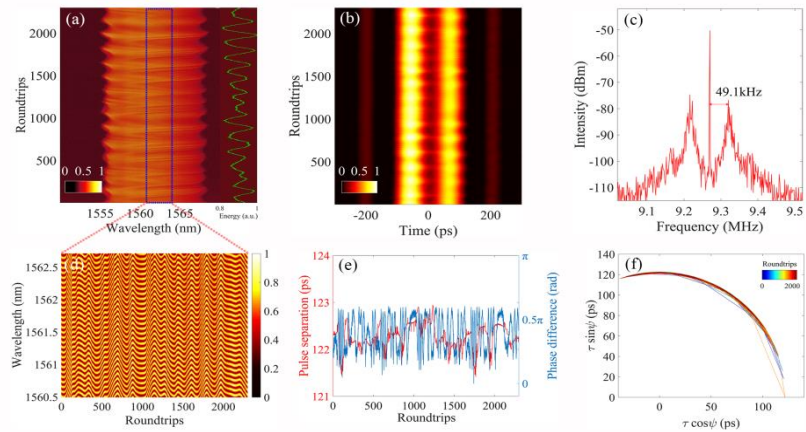

Figure 7. Characteristics of aperiodic pulsating soliton molecule. (a) The real-time spectra evolution measured with DFT, inset: the energy evolution. (b) The temporal evolution. (c) The RF spectrum. (d) The enlarged spectra. (e) The soliton separation and phase difference as a function of the roundtrips. (f) The corresponding interaction planes.

## 4. Discussions

Pulsating soliton is a distinct nonlinear local structure in dissipative systems. The investigations on both its temporal and spectral characteristics in real-time contribute to reveal the underlying abundant dynamics of pulsating solitons, which would also help to better design and optimize the laser for generating robust ultra-short pulse. In recent years, thanks to the development of the DFT technique, more and more transient dynamics of the pulses in fiber lasers have been unveiled [15-35]. They directly deepen our understanding of the process by which the pulses generated or evolved. In this work, we observed various dynamics of pulsating solitons by using DFT from explosive pulsating soliton to pulsating dual-soliton bunches and pulsating soliton molecules in a same fiber laser under engineering the pump power and the polarization state of the pulses. The explosive pulsating soliton events evolve from weak to strong, accompanying by the great broadening of the spectrum at the strongest explosive event. It may come from the fluctuation of pulse peak power. When the laser operates at pulsating regime, the gain and loss are not complete balance. Pulsating is an intermediate state between stationary stability and chaos. At certain time, the imperfect pulsating behavior leads to the perturbation of the pulse peak power. When the peak power of the mode-locked pulse exceeds the cavity tolerant nonlinear effect (overdriven nonlinear effect), the soliton will split into small pulses with random phase difference in this case and show the explosive behavior. After some roundtrips, some small pulses dissipate and the residual pulses reshape into one. Then the next perturbation of the pulse peak power will make the pulse split and the spectrum collapse again. The degree of the explosive events depends on the peak power of the split small pulses. The higher the peak power, the broader the spectrum generates due to the self-phase modulation and cross-phase modulation effects. And it is more likely to appear rogue waves as the statistical analysis.

When we increased the pump power and manipulated the orientation of the PC, the pulsating dual-soliton bunch with irregular or regular pulsating period were observed. Note that the separation between the two solitons in a bunch is more than 20 ns, they do not have direct interactions. Therefore, the pulsation for each individual could be irregular or regular, synchronous or asynchronous. In the case of the relatively stronger pulsation, the characteristics of their real-time spectra are similar with those of the pulsating single-soliton. However, for the pulsating soliton molecules, the separation between the two solitons inside the molecue is around 106 ps, with the soliton width of about 23 ps (See Figure S5 in Supplementary Materials). The two solitons are bounded together under their direct interactions due to the change of the pump power or/and the PC. Consequently, the spectrum presents periodical modulation in soliton molecules state. Generally, the two solitons inside the pulsating soliton molecules act as a unit. It is interesting to know why the pulsations of soliton molecules show different behaviors in the experiment. By comparing the two states of the soliton molecules, the periodic evolution of pulse separation and phase difference could be observed for the regularly periodic pulsation of soliton molecule. However, the irregularly periodic pulsation of soliton molecule shows the aperiodic evolution of pulse separation and phase difference. As we know, the soliton pulsation is related to the gain and loss in the laser cavity. In this case, the soliton molecule will possess the same state including soliton separation and phase difference when it propagates through the EDF every roundtrip for the periodically evolving phase difference and pulse separation. Thus, the gain for this type of soliton molecule will be equilibrated, resulting in the appearance of regular pulsation. However, the aperiodic evolution of soliton molecule will experience different states when passing through the EDF every roundtrip, leading to irregular gain dynamics of soliton molecule. Therefore, it is indicated that the pulsating behavior of soliton molecule is influenced greatly by the evolution of pulse separation and phase difference. That is to say, the locked or periodic evolution of pulse separation and phase difference is essential for achieving regular pulsation of soliton molecule. In fact, the soliton molecules with oscillating phase difference and separation have been experimentally observed [29, 30, 33-35], but there are rare reports on the simultaneous pulsation in energy. Furthermore, we also observed the pulsating triple-soliton molecule (See Supplementary Materials), indicating that the pulsating behavior of soliton molecule could be an intrinsic feature in ultrafast lasers. Although we have experimentally observed these diverse pulsating soliton patterns in a same fiber laser, it is anticipated to find more distinct pulsating solitons and reveal the detailed temporal information through using time-lens measurement technique [23,44,45].

## 5. Conclusion

To conclude, in this work, we have experimentally investigated the diversity dynamics of pulsating solitons from single-soliton to multi-soliton in a fiber laser by using DFT. The weak to strong explosive behaviors of pulsating soliton, as well as the rogue wave generation during explosions were observed. Moreover, we find that the pulsations of each individual in multi-soliton bunch could be regular or irregular. On the other hand, for the pulsating soliton molecule, the two solitons inside the molecule act as a unit, the simultaneous pulsations in energy, separation and relative phase difference were observed. These findings would further enrich the ultrafast dynamics of dissipative solitons in nonlinear optical systems.

**Acknowledgements**

We acknowledge the support from National Natural Science Foundation of China (Grant Nos. 61875058, 11874018, 61875242, 61378036), Science and Technology Program of Guangzhou (Grant No. 201607010245).

**Conflict of Interest**

The authors declare no conflict of interests.

# Supporting Information

## 1. The laser setup

The schematic of the laser setup is shown in Figure S1. The fiber laser cavity includes a 10 m long erbium-doped fiber (EDF) with dispersion parameter of −44.5 ps/nm/km pumped by a 980 nm laser diode through a wavelength-division-multiplexer (WDM), a polarization-dependent isolator (PD-ISO), a polarization controller (PC), a carbon nanotube saturable absorber (CNT-SA) and an output coupler (OC). The other fibers in the cavity are single-mode fibers (SMF-28) with 12 m long. Therefore, in this laser, the effect of saturable absorption is realized by both nonlinear polarization rotation (NPR) technique and CNT-SA. Furthermore, the output coupler with coupling ratio of 10:90 is used to extract 10% of the laser output for measurements.

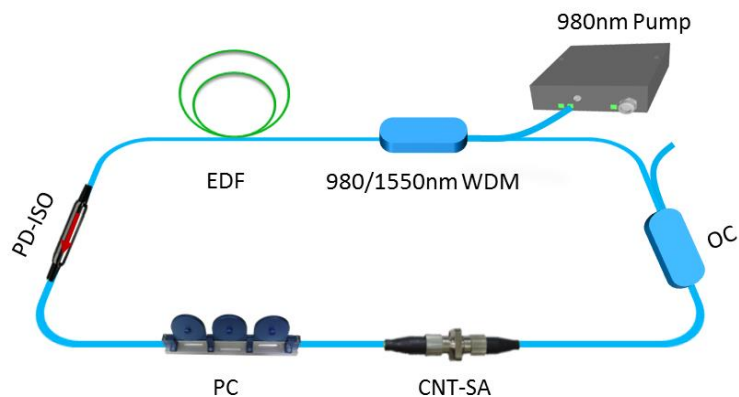

Figure S1. Schematic of the fiber laser.

## 2. Spectral variation of the pulsating single-soliton with the PC adjustment

Here, to better display the characteristics of time-average optical spectrum displayed on the OSA when adjusting the orientation of the PC, we provide several optical spectra corresponding to different PC orientations at the same pump power, as shown in Figures S2(a), (b) and (c). From these figures, one can see that by tuning the PC orientation, the shape of the optical spectrum varies in a certain range, from flat top to remarkable gradient edges in our experiment. Moreover, this kind of spectrum is connected with the pulsation parameters. To put it differently, the larger the gradient extent of the spectral edges, the stronger the modulation of pulsating soliton energy and amplitude is. The single-shot spectra corresponding to the maximum pulsating soliton energy for these three PC orientations through DFT are recorded in Figures S2(d), (e) and (f). As seen from Figures S2(d), (e) and (f), the single-shot spectra exhibit the structured profiles, where two sharp spectral peaks with oscillation structures around them locate at both edges of the spectrum. And the larger gradient degree of the average spectrum, the stronger the spectral fringe at the edges of the single-shot spectrum is. This structured profile of the single-shot spectrum is the characteristic of the dissipative soliton shaping in net-normal dispersion fiber laser [24]. As for the difference of the spectral fringe at the edges of the spectrum in these three cases, it comes from the increase of the soliton energy during pulsating. Therefore, the spectrum shows stronger oscillation structure at the edges.

Additionally, it should be noted that these three average spectra with diverse gradient extent edges correspond to the different pulsating solitons with different energy modulations, respectively. In other words, in our experiment, we find that the pulsating soliton energy modulation varies according to the gradient extent of average spectrum. When the energy modulation becomes larger, the structured spectrum corresponding to the maximum pulsating soliton energy varies, from peaks appearing at the single-shot spectral edges without remarkable fringes to sharp peaks with deep fringes.

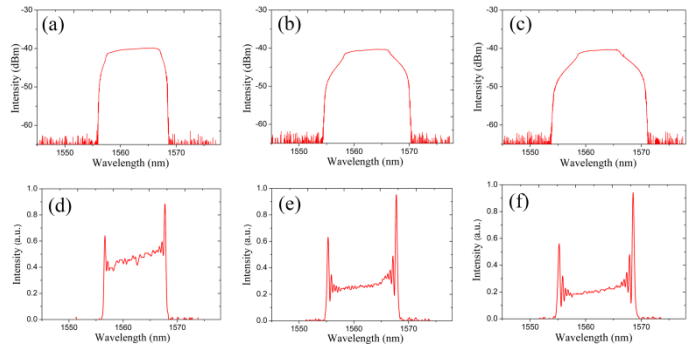

Figure S2. Optical spectra at different PC orientations. (a)-(c) Average spectra obtained from OSA. (d)-(f) Shot-to-shot spectra corresponding to the maximum pulsating soliton energy.

## 3. Rogue wave generation in explosive long-period pulsating soliton regime

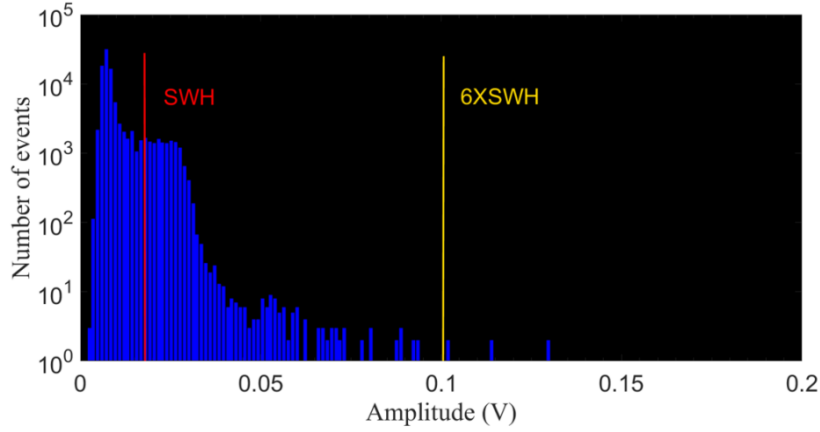

Figure S3. Statistical distribution histogram of the amplitude fluctuations in explosive long-period pulsating soliton regime.

Rogue wave (RW) has been observed in different operation regimes in fiber lasers, like in noise-like pulse and chaotic multi-pulse cases [R1-R3]. It seems that all these complex nonlinear phenomena to some extent share a common characteristic: solitons with chaotic behaviors. Since the explosive long-period pulsating soliton in our experiment has remarkable chaotic feature and the extreme spectral events appear, we perform statistical measurements to analyze the link between the highly chaotic pulsating soliton and generation of rogue wave. Note that, instead of studying the temporal solitons which are expected to generate the RW, due to our temporal resolution (80 ps) is not enough to resolve the timing between the splitting pulses from a soliton when explosion of pulsating solitons takes place, we investigate the extreme waves in the spectral domain by recording the maximum intensity of the single-shot spectrum. To display the histogram of the statistical distribution of the maximum spectrum intensity in the explosive pulsating soliton regime, 100000 trace samples are recorded by the oscilloscope. The results are shown in Figure S3. From Figure S3, we can see that the significant wave height (SWH) is 0.0171V. In these pulse amplitude events, the vast majority of them are concentrated in a lower intensity. The largest intensity of extreme spectral wave is more than 6 times of the SWH. Moreover, as shown in Figure 3(c), the chaotic spectral fringes happen around the maximum pulsating soliton energy. Therefore, we believe that RWs generated in such an explosive long-period pulsating soliton regime.

## 4. Typical single-shot spectrum of dual-soliton bunch with regular pulsation

To better show the characteristics of the single-shot spectrum of dual-soliton bunch with regular pulsation, three typical real-time spectra are taken within one pulsation period, as shown in Figure S4. One can observe that at the positions of the 335th and 365th roundtrips, the sharp peaks appear at the edges of the spectrum and gradually become stronger. However, at the 454th roundtrip, the spectrum has parabola-like top. These spectral characteristics are quite similar to those of pulsating single-soliton in the inset of Figure 2(c).

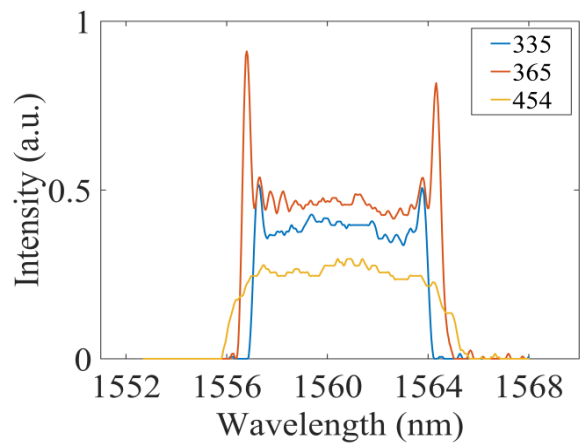

Figure S4. Single-shot spectra of dual-soliton bunch with regular pulsation.

## 5. The spectrum displayed on the OSA and the autocorrelation trace of the periodic pulsating soliton molecule

Figure S5 shows the spectrum presented on the OSA and the corresponding autocorrelation trace of the periodic pulsating soliton molecule. As can be seen from Figure S5(a), the spectrum shows evident fringes, which is the typical characteristic of a soliton molecule. The corresponding autocorrelation trace depicted in Figure S5(b) further confirms that the soliton molecule consists of 2 identical solitons. Moreover, the temporal separation of 106.5 ps coincides with the 0.08 nm modulation spacing on the spectrum.

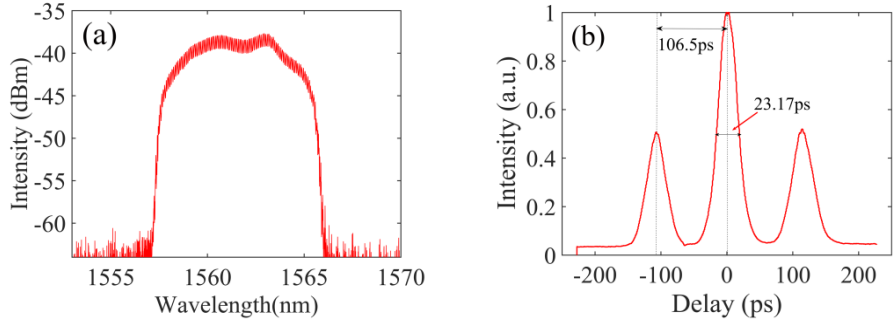

Figure S5. (a) The spectrum displayed on the OSA and (b) the autocorrelation trace of the periodic pulsating soliton molecule.

## 6. Characteristics of pulsating triple-soliton molecule

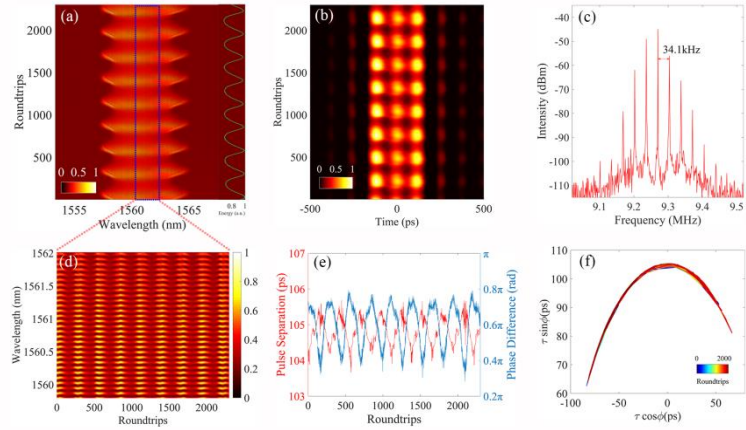

Figure S6. Characteristics of pulsating triple-soliton molecule. (a) The real-time spectra evolution measured with DFT, inset: the energy evolution. (b) The temporal evolution. (c) The RF spectrum. (d) The enlarged spectra. (e) The soliton separation and relative phase difference as a function of the roundtrips. (f) The corresponding interaction planes.

When increasing the pump power to 15.68 mW, we obtained pulsating triple-soliton molecule by carefully manipulating the PC. Figure S6 summarizes its characteristics. Figure S6(a) shows the spatial-spectral evolution. Obviously, the spectra with modulation pulsate periodically, like that of the dual-soliton molecule in Figure 6. For better clarity, we show the enlarged spectra in Figure S6(d). The energy evolution of the triple-soliton molecule is presented in the inset of Figure S6(a), which also varies periodically with a period of ~276 roundtrips, well consistent with the frequency difference of 34.1 kHz in RF spectrum in Figure S6(c). Furthermore, the corresponding temporal evolution of the triple-soliton molecule is demonstrated in Figure S6(b). It is clear that three pulsating solitons bound as a unit and evolves stably. In addition, we analyze the temporal separation and relative phase difference of the triple-soliton molecule, as shown in Figure S6(e). Note that the separation and phase difference between the solitons inside the molecule synchronously oscillate with the roundtrips, which are similar with those of the dual-soliton molecule in Figure 6(e). Figure S6(f) further shows the evolution trajectory of the interacting solitons in the phase plane, indicating that the oscillations possess relatively stable periodic behavior. So far, we observed the pulsating triple-soliton molecule, similar to the pulsating dual-soliton molecule, which pulsates not only in energy, but also in soliton separation and phase difference.